\documentclass[aps,prd,twocolumn, showpacs, groupedaddress]{revtex4}

\begin{document}
 

\title{Interactions and State Constraints via Induced Nonlinear Realizations of Lie Groups }

\author{Bill Dalton}
\affiliation{Department of Physics, Astronomy and Engineering Science 
St Cloud State University
}
\date{ \today}

\begin{abstract}
This is a study of induced nonlinear realizations of a Lie group G in which the presence of one field induces nonlinear transformations on another field.  The  covariant derivative structure is similar in form to that for local gauge theory.  For an arbitrary Lie group, basic equations and non standard invariant Lagrangian forms are described. Covariant constraint equations that place restrictions on field components are presented. With G = SU(2), a detail application to the electroweak model is discussed. We first show that the standard Lagrangian for the gauge electroweak model is invariant under these transformations. We then show that an alternate invariant Lagrangian is also possible. In it, the intermediate boson masses arise from the adjoint field rather than from  the Higgs doublet. An alternate invariant lepton Lagrangian is presented. A covariant constraint on the right-handed lepton field requires the right-handed neutrino field to vanish at the point where we obtain a massless (photon) field. Within this model, we have a clear explanation why weak interactions do not produce right-handed neutrinos. Neutrinos with and without mass are found. This model indicates an abundance of light mass leptons that are "blind" to the massless electromagnetic field but "see" four massive potentials. These more difficult to detect leptons would support a WIMP contribution to the missing mass.
\end{abstract}


\maketitle

\section{Introduction}
Directly confronting present theories are several experimental observations. We mention four. One of these is the complete absence of the right-handed neutrino in weak decay. A second is neutrino oscillations, an observation that can be described with neutrinos that have different masses. This requires an explanation for different neutrino masses together with difference in neutrinos as judged in the way they interact with other leptons. There is extensive evidence ( although indirectly ) of missing mass and missing energy in cosmological data. A fourth observation is that no free quarks are observed. 

Some of our "standard" models are constructed to accommodate, but not explain some of these observations such as no right-handed neutrino. The standard gauge models, in particular the electroweak model has been very successful in describing particle states and cross sections. Alternate, or extended models must include much of  the successful features of these models. 

This study is about a nonlinear transformation approach to interactions that has many of the successful features of gauge theory, while offering a plausible explanations for at least two of the above observations, no right-handed neutrino and neutrino mass. It raises possible explanations for other observations. It is a development of a limited version of the nonlinear realizations outlined in the Appendix of \cite{dc}. 

In these realizations, the nonlinear component of the transformation on a field $\Psi$ depends on the presence of one or more other fields which we call the inducing fields. The development here is for the case where one inducing field $V$ transforms via the adjoint representation. This correspond to linear transformations on a hypersphere with radius $V=\sqrt{V^kV^k}$ where the implied sum is over the number of generators of the group. For convenience we discuss how different fields, potentials and their features such as mass values are morphed from one point to another on this hypersphere.

A configuration points corresponds to a set $[\mathbf{L}, \mathbf{R},A_\mu, W_\mu^k, V^k]$ of left and right leptons, covariant potentials and inducing adjoint field components, all of which  can interact with other members at that point. Different points correspond in general to different physics. The symmetry transformations connect different configuration points of a given type. We classify points as disjoint if they cannot be connected by the symmetry. For example, points with different vector polarization modes are disjoint. Another example is points with different radius values $V$ on the hypersphere. We provide two non-standard invariants involving the potentials and $ V^k $  components for groups like $ SU(n) $ that have structure constants that are antisymmetric in all three indices. One of these invariants is completely algebraic and linear. The other involves a kinetic term. Configuration points that are not disjoint have the same value for these invariants. 

Any theory involving nonlinearity raises the question of superposition. Because the nonlinear transformations depend on the unit vectors $h^k = V^k/V$, superposition of potentials of two configuration points is allowed when the $V$ space components are parallel, that is, on the same $V$ space radius vector. Otherwise, superposition of potentials is generally not allowed. In the electroweak application, superposition of massless electromagnetic potentials is allowed because all massless potentials lie on the same radial vector. The allowed superposition of two parallel $V$ space vectors is very important because the net $V$ space radius scales the masses of the potential bosons. This brings up the possibility of coherence playing a possible role in enhancing the strength of the missing mass. 

Covariant constraint relation are described which, if imposed, can provide restrictive relations on the components of the field $\Psi$.  These covariant constraints do not restrict the transformations. Rather, they can forbid certain components of $\Psi$ from having nonzero values at certain configuration points while imposing relations between the components of $\Psi$ at other points. These constraints can offer possible explanations for certain unexplained physical observations such as the absence of the right-handed neutrino in weak decay. 

We give a detailed description for the electroweak model in this picture. First, we show that it is possible to reproduce the invariant Lagrangian structure of the standard gauge model.\cite{w} \cite{S}  We then provide an alternate Lagrangian structure. The same Yang-Mills \cite{ YM} structure is used for the potential kinetic part. For the mass terms for the potentials we use the new invariants available in these realizations. From these, the masses for the $Z_\mu$ and $W_\mu^{\pm}$ bosons arise from the adjoint field $V$. This is in sharp contrast to the gauge model where the Higgs doublet provides the masses. Using a right-handed lepton composed of a right-handed neutrino and right-handed electron spinor, is is easy to construct an invariant lepton Lagrangian. However, one is immediately confronted with the experimental data which does not support a right-handed neutrino in weak decay. This is one reason the Higgs doublet field was introduced. 

To confront this dilemma here, we use a covariant constraint on the right-handed lepton components. This constraint does not break the symmetry, it is covariant and thus holds at all points on the real hypersphere. Covariant constraint equations can be obtain from one of two eigenstates. The lepton physics completely differs for the two cases. At the pole $V_3=V$ (called the north pole here) the mass ratio $\frac{M_Z}{M_W} $  for the intermediate bosons take on the usual value, the mass for the $A_\mu$ field is zero and one constraint choice requires the right-handed neutrino to vanish, but only at this point. This is the only place on the hypersphere where this experimentally observed combination happens. At other places on the sphere the ratio $\frac{M_Z}{M_W} $ changes, the $A_\mu$ field becomes massive and the right-handed lepton is not zero. Leptons and potentials at points other than the north pole do not "see" the massless electromagnetic field. Instead, they see the heavy $A_\mu $ vector field. Different points on the $V$ sphere represent different physical states.  For instance, leptons can go from massive to massless at different points.  To change from one point to another in the laboratory requires a physical change of the $V^k$ fields. This is why we refer to these symmetries as active.

Only two points admit massless leptons, one at the north pole and one at the south pole $(V_3=-V)$. At all points on the sphere other than the north pole the neutrino has a mass. At all non pole points the $A_\mu$ field is massive and both leptons are massive. The massive leptons and four massive potentials in the non pole zone offer some support for a WIMP contribution for the missing mass problem.  This region corresponds to a "soup" of light mass leptons interacting with four massive vector bosons. At the south pole the left-handed electron spinor becomes massless and the right-handed electron spinor vanishes. In addition, the $A_\mu$ field is very massive at this pole. This "electron" component at this pole does not interact electromagnetically since there is no massless field there. These features lead to a question as to its identification and how could it be observed. It acts more like a neutrino than an electron.

This theory offers two separately conserved currents, one for the linear part and one for the nonlinear part. For the nonlinear part, the conserved current reduces to conservation of charge at the north pole. At other points where we do not have a massless $A_\mu$ field the interpretation is not clear.

Again, it is stressed that in this theory, physical interpretation of these symmetries is active. This means that points on the $V$ hypersphere corresponds to the physical presence of the corresponding component fields $V^k$. For the electroweak application a point on the sphere is determined by the presence of the $V_3$ field and the two $V^{\pm}$ fields. These field components help determine the nature of the boson fields as well as the lepton fields. There may be different $V$ space points and corresponding potentials at the same space-time point. Where superposition is permitted, the radius of the hypersphere could change. Since this $V$ sphere radius scales the masses, there is the clear possibility that coherence of the contributions to the $V^k$ fields could enhance the mass, and thus contribute to the missing mass. This raises a common question of concern. Is the field (whether the Higgs doublet, or the adjoint field) that scales the boson masses, a constant, or is it the net field resulting from a superposition that could differ in different parts of the universe? 

We treat the infinitesimal generators as differential operators in the group parameter space, with  $[T_a, GF] =[T_a, G]F + G[T_a, F] $. \cite{rac},\cite{ham} We indicate the group structure constants as $C^{abc}$.  Except for the space-time indices, we freely place the group indices as superscripts, or as subscripts as needed for simplicity. We use the space-time metric $g_{\mu\nu} = [1,-1,-1,-1]$. The realizations here are but one of many types of nonlinear realizations of Lie groups. A number of references to  similar realizations, including co-set realizations can be found in \cite{dc}, \cite{JP}, \cite{gs} and \cite{cs}.

In section \textbf{A} we discuss the infinitesimal transformations, covariant derivatives, and Yang-Mills fields for an arbitrary Lie group. The aim is to establish basic equations for applications for, and beyond, the $SU(2)$ electroweak application here. In  section \textbf{B} we discuss superposition, covariant constraints and two invariants for a Lie group in which the structure constants are totally antisymmetric in the three indices. These invariants can be used to construct invariant Lagrangians. In section \textbf{C} we show that the standard Lagrangian for the gauge electroweak model with the Higgs doublet is invariant with these transformations. In section \textbf{D} we use the new invariants and covariant constraints to construct an alternate Lagrangian for the electroweak interaction. Finally we discuss in detail the two conserved currents in the \textbf{Appendix}. 

\subsection{ Infinitesimal Transformations}
We consider the Lie algebra of a continuous symmetry group $G_1$ acting simultaneously on a field $\Psi$ and one or more fields represented here by  $\Phi$. In one special category, these infinitesimal transformations are expressed in part in terms of generators of a second continuous group $ G_2$. We refer to $G_2$ as the  "hidden" symmetry group. These realizations follow the outline given in the appendix of \cite{dc} . In general, for $g(\delta\alpha)\in  G_1 $, where $\delta\alpha $ represents the infinitesimal group parameters, we study the simultaneous transformations
 \begin{equation}\label{A:1}
g(\delta\alpha) \colon 
 \begin{cases}
 \Phi  \to  \Phi^{\prime}(\delta\alpha,\Phi),\ 
 \Psi   \to  \Psi^{\prime}(\delta\alpha,\Phi,\Psi)
 \end{cases} 
\end{equation}
The particular feature in equation (\ref{A:1}) to notice here is that the transformation on the field $\Psi$ depends on the field $ \Phi$. The first order expansions of these transformations are written as follows.
\begin{equation}\label{A:2}
\Psi^\prime  = U \Psi= \Psi + \delta\alpha^a[T^a,\Psi]
\end{equation}
\begin{equation}\label{A:3}
\Phi^\prime  = \Phi+\delta\alpha^a[T^a,\Phi]
\end{equation}
We use the convention that  the indices $a,b..$ label the generators and components of $G_1$ and that a repeated index implies summation over the $n_1$ generators of the algebra.  If the Lie algebra is to be satisfied when acting on a function $ S $, the generators  $T^a$ must satisfy the commutation rules;
\begin{equation}\label{A:4}
[T^a,[T^b,S]]-[T^b,[T^a,S]]=C^{abc}[T^c,S]
\end{equation}
The associative property of $G_1$ require the structure constants  $C^{abc}=-C^{bac} $ to satisfy the Jacobi identity.
\begin{equation}\label{A:5}
C^{abc}C^{cdm}+C^{bdc}C^{cam}+C^{dac}C^{cbm}=0
\end{equation}
 Acting on an $N$ component field $\Psi$, the infinitesimal generator action studied in this paper has the following form. 
  \begin{eqnarray}\label{AA:6}
[T^a,\Psi] =    
  (i\xi^a\frac{1}{2} Y -  h^a)\Psi    
\end{eqnarray}
Here, the operator $Y$ is proportional to the unit matrix with $Y\Psi=y\Psi$. The group parameters are global, but the local nature of the transformations arises via the dependence of the transformations on the fields $\xi$ and $h$. This is the major difference between these transformations and those of local gauge theory. In physical interpretation, the $\xi$ and $h$ fields must be considered in conjunction with the covariant potentials required because of the local nature of the transformations. For the special case where the transformations induce a group $G_2$, we have
 \begin{eqnarray}\label{A:6}
[T^a,\Psi] =    
  (i\xi^a\frac{1}{2} Y - h^a_i\tau^i )\Psi .   
\end{eqnarray}
The  $ \tau^i $  are $ N \times N $ matrices that generate a linear representation of the group $ G_2 $ and are chosen to satisfy the following relation.
\begin{equation}\label{A:7}
[\tau^i,\tau^j]= -C^{ijk}\tau^k 
\end{equation}
We use the convention that  the indices $i,j..$ label the generators and components of $G_2$ and that a repeated index implies summation over the $n_2$ generators of the algebra of $G_2$.   The following general relation follows from $(\ref{A:4})$.
\begin{eqnarray}\label{A:8}
\frac{i }{2}([T^a,\xi^b] - [T^b,\xi^a] - C^{abc}\xi^c) Y\Psi  \nonumber \\ -([T^a,h^b ] -  [T^b,h^a]  - [h^b,h^a] - C^{abc}h^c )\Psi = 0
\end{eqnarray}
We consider the case where this equation splits into two equations.
\begin{equation}\label{AA:8}
[T^a,\xi^b] - [T^b,\xi^a] - C^{abc}\xi^c_k = 0
\end{equation}
\begin{equation}\label{A:9}
[T^a,h^b ] -  [T^b,h^a]  -  [h^b,h^a] - C^{abc}h^c =0
\end{equation}
For the special case described by (\ref{A:6}) equation (\ref{A:9}) becomes
\begin{equation}\label{AA:9}
[T_a,h^k_b ] -  [T_b,h^k_a]  +   h^i_b h^j_aC^{ijk} -  C^{abc}h^k_c=0.
\end{equation}
For arbitrary $G_1$, the number of commutation equations compared with the number of field variables can easily lead to over conditioning. In (\ref{AA:8}) there are $\frac{n_1(n_1-1)}{2} $ equations but only $n_1$ components $\xi^a$. The number of equations exceeds the number of field variables for $n_1 > 3$. Even so, realizations that satisfy the constraint (\ref{AA:8}) can be found. For instance, many of the nonlinear realizations of $SL(2,C)$ in \cite{JP} satisfy this condition, but the adjoint realization does not. For $G_1=SU(2)$ for example there are three $\xi^a$ components appearing in (\ref{AA:8}). The generator action for linear transformations on a fundamental doublet in matrix form is 
\begin{equation}
[T_i, \xi] =\tau_i\xi, \hspace{.3cm}\xi= \left(\begin{array}{c}\xi_3+i\xi_4 \\\xi_1+i\xi_2\end{array}\right)\hspace{.3cm} \tau_i=\frac{i}{2}\sigma_i .
\end{equation}
The structure constants are the Levi-Civita tensors $C^{ijk}=\epsilon^{ijk}$. As operators acting on the individual components we have 
\begin{equation}\label{WNL}
[T_i,\xi_j] =\frac{1}{2}[-\xi_4\delta_{ij} +\epsilon_{ijk}\xi_k], \hspace{.3cm} [T_i,\xi_4] = \frac{1}{2}\xi_i.
\end{equation}
These relations satisfy (\ref{AA:8}), but there may be others that also satisfy (\ref{AA:8}).

To obtain covariant field equations for the local transformations in (\ref{AA:6}), covariant derivatives are defined as in local gauge theory. 
\begin{equation}\label{A:12}
D_\mu \Psi =   \partial_\mu\Psi +  i d \frac{1}{2} B_\mu Y  \Psi  - \gamma  W_\mu \Psi  
\end{equation}  
\begin{equation}\label{A:13}
(D_\mu \Psi)' = UD_\mu \Psi 
\end{equation}
With $R_\mu = i d \frac{1}{2} B_\mu Y - \gamma  W_\mu $ equation (\ref{A:13}) gives the following standard relation.
\begin{equation}\label{A:14}
R_\mu' = U R_\mu U^{-1} - (\partial_\mu U)U^{-1}
\end{equation}
Here, $B_\mu Y$ and  $W_\mu$ are the diagonal and nondiagonal field potentials with respective coupling constants $d$ and $\gamma$.  The field tensors given in terms of $R$ are defined as in Yang-Mills gauge theory as follows.
\begin{equation}\label{A:15}
R_{\mu\nu} = \partial_\mu R_\nu - \partial_\nu R_\mu +   [R_\mu, R_\nu]
\end{equation}
With (\ref{A:12}) the transformation of these tensors is
\begin{equation}\label{A:16}
R_{\mu \nu}' = U R_{\mu\nu} U^{-1}.
\end{equation}
The above covariant relations are identical in form to those of  standard Yang-Mills gauge symmetries except here $R_\mu $ contains both the diagonal and non diagonal terms. Transformations of the diagonal and non-diagonal components are considered below. Because of the presence of the fields $\xi$ and $ h$ we have some differences in detail. Because $Y$ is proportional to the unit matrix, we can satisfy the infinitesimal form of (\ref{A:14}) with the following relations.
\begin{equation}\label{A:17}
[T^a,B_\mu] = - \frac{1}{d} \partial_\mu\xi^a
\end{equation}
\begin{equation}\label{A:18} 
[T^a, W_\mu]  = -[h^a,W_\mu] - \frac{1}{\gamma} \partial_\mu  h^a
\end{equation} 
With $B_{\mu\nu}=\partial_\mu B_\nu  - \partial_\nu B_\mu$ and 
\begin{equation}\label{YFT}
W_{\mu\nu}=\partial_\mu W_\nu  - \partial_\nu W_\mu-\gamma [W_\mu, W_\nu]  
\end{equation}
we have 
\begin{eqnarray}
W^{\prime}_{\mu\nu}  = U W_{\mu\nu} U^{-1},   B^{\prime}_{\mu\nu}  = B_{\mu\nu} . 
\end{eqnarray}
For the special case (\ref{A:6}), we assume  $W_\mu = W_\mu^j \tau^j $ giving
\begin{equation}\label{A:19} 
[T^a, W_\mu^l]  = -   h^a_i C^{ikl}_mW_\mu^k - \frac{1}{\gamma} \partial_\mu  h^a_l.
\end{equation}
This expression involves the potentials as well as the field $h$.
The fields $\xi$ and $h$ may or may not appear explicitly in the Lagrangian. However, conservation rules corresponding to the  $G_1$ symmetries will involve the $\xi$ and $h$ fields. Let $F_i$ represent all fields, including potentials, involved in a Lagrangian and that satisfy the Euler-Lagrange field equations. For each parameter of $G_1$ Noether's conservation theorem reads.
 \begin{equation}\label{cc} 
\partial_\mu (\frac{\partial K}{\partial F_{i,\mu}} [T^a,F_i])=\partial_\mu J_a^\mu = 0
\end{equation} 
The $[T^a,F_i]$ factors in the current components depend on the fields $\xi$ and $h$. The conserved current for a particular nonlinear realization of the Lorentz group was described in detail in \cite{dc}. There, it was shown that the nonlinear current component was separately conserved.

\subsection{Invariant Forms, Superposition and Covariant  Constraint}

The covariant potentials combined with the inducing fields $\xi^a$ and $h^a$ offer a variety of possibilities in modeling physics. Here, we discuss a basic nonlinear realization type that is extended from a linear realization. We then consider some invariant forms. These can be used to classify disjoint configurations which in this model correspond to different physical states.  We then discuss covariant constraints that if invoked prevents certain components of $\Psi$ from existing in certain limits. One constraint will be used in a latter section to provide an explanation in this model of the absence of the right-handed neutrino in observed weak interactions. The  discussion in this section is for arbitrary groups, but will be used with $G_2=SU(2)$ in the next section where we discuss the electroweak model in this picture. First, consider the case where $G_1=G_2$ with 
\begin{equation}\label{VV:1}
h^a_j = -\delta_j^a + Z^a_j .
\end{equation}
If the $Z^a_j$ and $\xi$ vanish, these realizations reduce to the usual linear representations of $G_1$. Put another way, the usual linear realizations of a Lie group can be extended to this type of nonlinear realization by including the $\xi$ and $Z$ fields. We refer to this type of nonlinear realization as extended. This type of extension for the diagonal term only was studied in detail for the group $SL(2,C)$ in  \cite{dc}.  The constraint  (\ref{AA:9}) becomes
\begin{eqnarray}
[T_a,Z_b^k] - [T_b,Z_a^k]  + Z_a^jC^{jbk}  - Z_b^iC^{iak} \nonumber \\  + Z_b^i Z_a^j C^{ijk} = C^{abc}Z_c^k.
\end{eqnarray}
This equation is satisfied with the relations
\begin{equation}\label{VT}
Z_b^k= -Z_b h^k, \hspace{.2cm} h^2=h^kh^k=1,\hspace{.2cm} [T^a,h^k] =    C^{aki}h^i   
\end{equation}
The $h^k$ components transform via the adjoint representation and the $Z_b$ satisfy 
\begin{equation}
[T^a,Z^b] - [T^b,Z^a] - C^{abc}Z^c_k = 0.
\end{equation}
This equation is identical to (\ref{AA:8}) so that the $Z_a$ could be identified with the $\xi_a$ as far as the group action is concerned, but we emphasize that this is just one choice.
The potentials transform as follows.
\begin{eqnarray}
[T_a,W_\mu^l]=  C^{alk}W_\mu^k - Z_ah^iC^{ikl}W_\mu^k + \frac{1}{\gamma}\partial_\mu(Z_ah^l)
\end{eqnarray}
From this expression, it is clear that the nonlinear second term forbids superposition of the potentials $W_\mu^l$ for configuration points that have different $h^k$ components. The $h^k$ represent the unit vector components in the hypersphere.This means that superposition is allowed only for points in the $V$ hypersphere that lie along the same radial direction. 
With $V^k =Vh^k$ where $V$ is a group invariant, and with $Z_a= \xi_a$, consider the composite boson field $F$ with components defined as
\begin{eqnarray}
F_\mu= \gamma W_\mu^l V^l+\beta_\mu V d. 
\end{eqnarray}
For  a $ G_1$  such as $SU(n)$ with structure constants antisymmetric in all three indices, it is easy to show 
$[T_a, F_\mu]=0 $ so that the $F_\mu$ are invariant under $G_1$. We have the following form that is invariant under both $G_1$ and the Lorentz group. 
\begin{eqnarray}\label{Kb}
K_b = \frac{1}{2} F_\mu F^\mu 
\end{eqnarray}
This invariant is independent of the eigenvalue of the diagonal operator $Y$ and depends only on the potentials and the $h^k$ components and two coupling constants.

We can construct a second invariant using the following quantities.
\begin{equation}
C_\mu^l = V^i\epsilon^{ilk}\partial_\mu V^k.
\end{equation}
With these quantities consider the following expression.
\begin{eqnarray}\label{Ka}
K_a = \big[\frac{\gamma^2}{2}\big(V^2W^l_\mu W_l^\mu - W_\mu^l V^lW^\mu_k V^k \big) + \frac{\gamma}{a}W_\mu^lC_l^\mu \big] 
\end{eqnarray}
The constant $a$ in the last term is needed to accommodate the normalization of the structure constants which appear in this term via the $C_\mu^k$.  A bit of work produces the relation
\begin{eqnarray}
[T_b,K_a] = \gamma V^2  \xi_b \Big[ W_\mu^l\partial_\mu h_l \nonumber \\ +\frac{1}{a}(C^{ikl} C^{lsn} h_ih_n)W_\mu^k\partial^\mu h_s \Big].
\end{eqnarray}
One can show that the first bracket factor of the second term is symmetric in the indices ($k,s$) which we write as follows.
\begin{eqnarray}\label{SN}
C^{ikl} C^{lsn} h_ih_n = b(\delta_{ks} -h_kh_s)
\end{eqnarray}
The constant $b$ depends on the normalization of the structure constants for the group $G_1$, and must be calculated from this equation once the basis for the structure constants has been chosen. For $SU(2)$ with $C^{ijk} = \epsilon^{ijl}$ one can easily show $b=-1$. With the choice $a = b$ and $h^k h^k=1$ the second term in (\ref{SN}) when used in (\ref{Ka}) will give zero and we immediately obtain $[T_b,K_a] =0$ so that $K_a$ is an invariant under the group. By structure, it is also invariant under the Lorentz group. Since the last term in (\ref{Ka}) has the $a^{-1}$ factor, this form is invariant for any choice of normalization of the structure constants. 

With  $Z_a= \xi_a$ the transformation on $\Psi$ becomes
\begin{equation}\label{TS}
[T_a,\Psi] = \tau_a\Psi +i\xi_a\frac{1}{2}[Y \mathcal{U}+H ]\Psi, \hspace{.3cm} H = -2 i  h^k \tau^k 
\end{equation}
where $\mathcal{U}$ is the unit matrix. The number of $h^k$ components is equal to the number of group generators.

To form a quadratic invariant form $\tilde{\Psi} \Psi $, we define  $\tilde{\Psi} $  to transform as
\begin{equation}\label{V:1}
[T^a,\tilde{\Psi}] =  - \tilde{\Psi} (i \xi^a\frac{1}{2} Y -  h^a )   
\end{equation}
The form $\tilde{X}Z$ here is shorthand for the group scalar product $(X,Z)$ as defined in \cite{ham}.
Consider a second field $\Phi$ that transforms like $\Psi$.
 \begin{equation}\label{V:2}
[T^a,\Phi] =    
  (i\xi^a\frac{1}{2} Y - h^a )\Phi    
\end{equation}
 From these relations it immediately follows that the forms $\tilde{\Phi}\Psi$ and  $\tilde{\Phi} \Phi$ are invariant. 
Consider the matrix eigenvalue equation for the matrix $H$ 
\begin{eqnarray}\label{CG}
H\Psi=\lambda\Psi.
\end{eqnarray}
For  a $ G_1$  such as $SU(n)$ with structure constants antisymmetric in all three indices, this relation is covariant. For each eigenvalue $\lambda$, it represents a set of $N$ constraint equations involving the $N$ components of $\Psi$ and the $n_1$ components of the $h$ field, but does not involve the potentials $B_\mu$,$W_\mu^l$. We emphasize that the number of constraint equations in the set grows with the representation size $N$. The transformations is linear on the $h$ space hypersphere with unit radius $h^k h^k=1$.  It is convenient to view how the constraints on the components of $\Psi $ imposed by  (\ref{CG}) change with positions on this hypersphere. 

The invariant expressions $K_a$ and $K_b$ above involves the potentials $B_\mu$ , $W_\mu^l$ and the components $h^k$, but not components of  $\Psi$. In addition, the invariance of the $F_\mu$ under $G_1$ provides a linear relationship between the potentials at different points on the hypersphere. 

The relations of the potentials on the hypersphere taken together with a constraint  like (\ref{CG}) on the components of $\Psi$ offers a possible means of explaining certain experimental observations. For example, for the electroweak application in a later section we show that the vector potential $A_\mu$ becomes massless at the north pole ($h_3=1$) on the unit sphere. The constraint (\ref{CG}) used on the right lepton field mandates that the right neutrino $\nu_R$  must vanish at this pole point. This is one plausible explanation of one experimental observation. Viewed differently, we could say that the massless photon is confined to the north pole and the right neutrino is confined elsewhere on the sphere. This is a different view of confinement. This immediately raises the question as to whether a constraint like (\ref{CG}) together with the potential relations on a hypersphere for $G_1 = SU(3)$ could offer an explanation for quark confinement. 

Examination of the invariant forms $K_a$ and $K_b$ shows that the mass terms for the potentials take on different values on the hypersphere, so that the interpretation of the particle will depend on the position on this sphere. For example, the mass term for the $A_\mu$ field just mentioned only vanishes at the north pole, so that only at this location can we view this field as the electromagnetic vector potential. At other points, it is a massive vector field with significant interactions with other potential fields.

 The above realizations offers a means of constructing kinetic invariants other than the standard Yang-Mills form. Since the $F_\mu$ are invariant under $G_1$, and the $V^k=Vh^k$ transform linearly, we immediately have two  invariant forms.
 \begin{equation}
K_{FL} = \partial^\mu F_\rho \partial_\mu F^\rho, \hspace{.2cm} K_{VL} = \partial^\mu V_k \partial_\mu V^k,
\end{equation}
In addition, any function of of the invariant $V^2 = V^kV^k$ can be used.

\subsection{ Standard Lagrangian with Nonlinear $SU(2)$ }

In this section, we consider the basic electroweak picture within this framework with $G_1 = SU(2)$. Using a Higgs doublet,  and $Z_a = \xi_a$ we arrive at a Lagrangian that is basically the same as in the standard gauge picture  \cite{ w}, \cite{S}.  The difference between the approaches occurs in the conserved currents. These separate into a linear and nonlinear part, each of which is separately conserved. The nonlinear part involves the $\xi_a$ and $h^k$ quantities. Because of the involved detail, we relegate the current development to an appendix. In the following section, we describe an alternate Lagrangian structure. Below we provide some detail to facilitate comparison of the two Lagrangians.

We take $G_2=SU(2)$ with $ C^{ijk} = \epsilon^{ijk} $ with $\epsilon^{111} = 1$, and we set  $\tau^k=\frac{i}{2}\sigma^k$. The usual weak hypercharge eigenvalues are assumed.  Following the gauge electroweak model the left spinor $SU(2)$ doublet state is defined as follows.
\begin{equation}
\mathbf{L} = \left(\begin{array}{c}\nu_L\\e_L\end{array}\right),   \tilde{\mathbf{L}} = \left(\begin{array}{cc} \bar{\nu_L}& \bar{e_L}\end{array}\right) 
\end{equation}
Here, we use bold notation for the an eight dimension vectors made up of a stacked of two four dimensional spinors. To construct the Lagrangians, the following definitions are useful.
\begin{eqnarray}
\mathbf{ \mathrm{I}} = \left(\begin{array}{cc}\mathcal{U} & 0 \\ 0 & \mathcal{U}\end{array}\right), 
\mathbf{ \Gamma_\mu} = \left(\begin{array}{cc} \gamma_\mu& 0 \\ 0 & \gamma_\mu\end{array}\right), 
\mathbf{\sigma_3} = \left(\begin{array}{cc}\mathcal{U} & 0 \\ 0 & -\mathcal{U}\end{array}\right) \nonumber \\  \mathbf{\sigma_2} = \left(\begin{array}{cc} 0& -i\mathcal{U} \\ i\mathcal{I} & 0\end{array}\right), \mathbf{\sigma_1} = \left(\begin{array}{cc} 0& \mathcal{U} \\ \mathcal{U} & 0\end{array}\right) \hspace{2cm}
\end{eqnarray}
Here, $\mathcal{U}$ is the unit matrix in four dimensions. In this section we use the usual eigenvalues for the electro-weak hyperfine operator, $Y\mathbf{L}=-1\mathbf{L}$ and $Ye_R=-2e_R$. (This is changed in the following section.) We have the following group action on the $h^k$ and on $\mathbf{L}$.
\begin{eqnarray}\label{Tspn}
[T_a,h^k] = \epsilon^{akl}h^l, \hspace{.5cm} h^{\pm} = \frac{1}{\sqrt{2}}(h_1 \pm ih_2) \hspace{.7cm} \\\
[T_a,\mathbf{L}]     =  \frac{i}{2}\mathbf{\sigma_a} \mathbf{L} + i\xi_a \mathbf{\mathcal{T^-} }\mathbf{L}, \hspace{.3cm} \mathbf{\mathcal{T^-}} = \frac{1}{2}\big(-\mathbf{I}+ \mathbf{H}\big)\nonumber \\\ \mathbf{H} = \left(\begin{array}{cc}h_3 \mathcal{U} & \sqrt{2}h^-\mathcal{U} \\  \sqrt{2}h^+\mathcal{U} &  -h_3\mathcal{U} \end{array}\right)\mathbf{L}\hspace{2.5cm}  
\end{eqnarray}
For the singlet component we have
\begin{equation}
[T_a, e_R] = \frac{i}{2} \xi_a Y e_R = -i\xi_a e_R .
\end{equation}
 We use the following $SU(2)$ doublet, Lorentz scalar fields with $Y=+1$.
\begin{equation}
 \Phi =\left(\begin{array}{c}\Phi_1 \\\Phi_2\end{array}\right), \nonumber \\  \Psi =\left(\begin{array}{c}\bar{e_R}\nu_L \\\bar{e_R}e_L\end{array}\right)
\end{equation}
The generator action for $T_a$  on $\Phi$, or $\Psi$ is expressed as
\begin{eqnarray}\label{Tsca}
[T_a,\Phi] =  \frac{i}{2}\sigma_a\Phi +   i\xi_a\mathbf{\mathcal{T^+}}\Phi , \hspace{.5cm} \mathbf{\mathcal{T}}^+ = \frac{1}{2}(\mathcal{U}   + H), \nonumber \\\      
H = \left[\begin{array}{cc}h_3 & \sqrt{2}h^- \\  \sqrt{2}h^+ & -h_3 \end{array}\right]  \hspace{.2cm}
\end{eqnarray}
Here, the actions on $\mathbf{L}$,  $\Phi$ and $\Psi$ have  a linear and nonlinear component, where the latter involve the $\xi^a$ and $h^k$ fields.
We have the following invariants under both $G_1$ and the Lorentz group.
\begin{equation}  
\Phi^\dag\Phi,    \Psi^\dag\Psi,    \Phi^\dag\Psi,    \Psi^\dag\Phi
\end{equation}
Using  $\mathcal{I=\left(\begin{array}{cccc}1 & 1 & 1 &1 \end{array}\right)}$ and 
\begin{eqnarray}
\mathbf{\Phi}=\left(\begin{array}{c}\Phi_1 \mathcal{I} \\ \Phi_2 \mathcal{I}\end{array}\right), \mathbf{\tilde{\Phi}} = \mathbf{\Phi^\dag},
\end{eqnarray}
we have the relations
\begin{equation} 
 \Psi^\dag \Phi = \Phi_1 \bar{\nu_L} e_R + \Phi_2  \bar{e_L} e_R = \tilde{\mathbf{L}}\mathbf{\Phi}e_R
\end{equation}
\begin{equation}
 \Phi^\dag \Psi=( \Psi^\dag \Phi )^\dag=\Phi_1^\ast \bar{e_R}\nu_L+\Phi_2^\ast \bar{e_R}e_L = (\tilde{\mathbf{L}}\mathbf{\Phi}e_R)^\dag .
\end{equation}
To guide our analysis, consider first the following invariant  that is a common component  in the Lagrangian for the electroweak model.
\begin{eqnarray}\label{K3}
K_3  = -G_e( \Psi^\dag \Phi  +  \Phi^\dag \Psi) \nonumber \\  =  -G_e( \Phi_1 \bar{\nu_L} e_R + \Phi_2  \bar{e_L} e_R +\Phi_1^\ast \bar{e_R}\nu_L+\Phi_2^\ast \bar{e_R}e_L) \nonumber \\ = -G_e(\Phi_1 \bar{\nu_L} e_R + \Phi_1^\ast \bar{e_R}\nu_L)  \nonumber \\ +G_e \frac{1}{2}(\Phi_2-\Phi^\ast_2) (\bar{e_R}e_L -  \bar{e_L} e_R )  \nonumber \\  -G_e\frac{1}{2}(\Phi_2+\Phi^\ast_2) (\bar{e_R}e_L +  \bar{e_L} e_R )
\end{eqnarray} 
In the standard electroweak model the last term in this expression becomes, in a limit, the electron mass times $\bar{e}e $  where $e = e_L+e_R$ is the electron spinor. To describe the electron's mass,  $\Phi_2$ must become in some limit, a real constant commonly indicated by $\frac{\nu_o}{\sqrt{2}}$. 

With the parameter notation change  $(d,\gamma) \to (-g^\prime, g) $, the transformations on the potentials are given by
\begin{equation}\label{Tbp}  
[T_a,B_\mu] = \frac{1}{g^{\prime}}\partial_\mu \xi_a  ,
\end{equation}
\begin{equation}\label{Twp}
[T_a, W_\mu^l]  =   \epsilon^{alk} W^k_\mu - \xi_a h^i \epsilon^{ikl} W_\mu^k +\frac{1}{g}\partial_\mu ( \xi_a h^l).
\end{equation}
From these we have
\begin{eqnarray}
[T_a,W_\mu^3] = \epsilon^{a3k}W_\mu^k -\xi_a[h^1W^2_\mu -h^2W^1_\mu]  \nonumber \\ + \frac{1}{g}\partial_\mu(\xi_ah^3) 
= \epsilon^{a3k}W_\mu^k - i\xi_a[h^+W^-_\mu -h^-W^+_\mu]  \nonumber \\ +\frac{1}{g}\partial_\mu(\xi_ah^3), \hspace{1cm} 
W_\mu^{\pm} = \frac{1}{\sqrt{2}}(W_\mu^1\pm i W_\mu^2)
\end{eqnarray}
\begin{eqnarray}
[T_a,W_\mu^{\pm}] = \frac{1}{\sqrt{2}}\big(\epsilon^{a1k} \pm i \epsilon^{a2k}\big)W^k_\mu  \pm i\xi_a h^{\pm}W_\mu^3 \nonumber \\ \mp i \xi_aW_\mu^{\pm}h^3 +\frac{1}{g}\partial_\mu(\xi_ah^{\pm}).
\end{eqnarray}
Here, the action on the $W^l_\mu$  potentials has a linear and nonlinear component, where the latter involve the $\xi^a$ and $h^k$ fields.

Even though the group action on the potentials depends on the fields $\xi^a$ and $h^k$,  the covariant derivatives acting on $\Phi,\mathbf{L}  $ and $ e_R$ have the same general forms as in the common electroweak gauge model. With the above notation, we have the following covariant derivative forms.
\begin{equation}\label{D:2}
D_\mu \mathbf{L} = \partial_\mu \mathbf{L}   - \frac{i}{2}\mathbf{P}\mathbf{L} 
\end{equation}
\begin{eqnarray}\label{P}
\mathbf{P} = \left[\begin{array}{cc} (g W^3_\mu-g^\prime \beta_\mu) \mathcal{U} & g (W^1_\mu-i W^2_\mu) \mathcal{U} \\ g (W^1_\mu+i W^2_\mu) \mathcal{U} &( -g W^3_\mu-g^\prime  \beta_\mu) \mathcal{U} \end{array}\right]  \nonumber \\
=  \left[\begin{array}{cc} N Z_\mu \mathcal{U} & g\sqrt{2}W^- \mathcal{U} \\ g\sqrt{2}W^+ \mathcal{U} &[ -N\cos{2\theta_w} Z_\mu-2qA_\mu] \mathcal{U} \end{array}\right] 
\end{eqnarray}
The charge is given by 
\begin{equation}
q = \frac{g^{\prime} g}{N}.
\end{equation}
\begin{eqnarray}\label{D:3}
D_\mu e_R = \partial_\mu e_R +  iB_\mu g^{\prime} e_R \nonumber \\
=\partial_\mu e_R   -iN\sin{\theta_w}^2 Z_\mu  e_R +iqA_\mu e_R
\end{eqnarray} 
\begin{equation}\label{DS:2}
D_\mu \Phi = \partial_\mu \Phi- \frac{i}{2} X_\mu \Phi 
\end{equation}
\begin{eqnarray}\label{X}
 X_\mu =  \left[\begin{array}{cc} g W^3_\mu+g^\prime \beta_\mu & g\sqrt{2}W^- \\ g\sqrt{2}W^+& -g W^3_\mu+g^\prime \beta_\mu \end{array}\right] \nonumber \\\ = \left[\begin{array}{cc}[ N\cos{2\theta_w} Z_\mu + 2qA_\mu] &  g\sqrt{2}W^- \\g\sqrt{2}W^+ & -N Z_\mu\end{array}\right] 
\end{eqnarray}
In the above we used the standard potential relations.
\begin{equation}
\left(\begin{array}{c}W_\mu^3 \\\beta_\mu\end{array}\right)=\left(\begin{array}{cc}cos(\theta_w) & sin(\theta_w) \\-sin(\theta_w) & cos(\theta_w)\end{array}\right) \left(\begin{array}{c}Z_\mu \\A_\mu\end{array}\right)
\end{equation}
\begin{equation}
\cos(\theta_w) = \frac{g}{N}, \sin(\theta_w)= \frac{g^{\prime}}{N}, N = \sqrt{(g^{\prime})^2 + g^2}
\end{equation}

To facilitate constructing the conserved currents, we use a real Lagrangian. For the spinor fields we have the following real invariants.
\begin{eqnarray}\label{K1}
K_1= \frac{1}{2}[i  \tilde{\mathbf{L}}\mathbf{\Gamma}^\mu D_\mu \mathbf{L}  +(i  \tilde{\mathbf{L}}\mathbf{\Gamma}^\mu D_\mu \mathbf{L} )^\ast] \nonumber \\ =\frac{i}{2}[\bar{\nu}_L\gamma^\mu( \partial_\mu \nu_L)-(\partial_\mu \bar{\nu_L})\gamma^\mu \nu_L]  \nonumber \\ +\bar{e}_L\gamma^\mu( \partial_\mu e_L)-(\partial_\mu \bar{e_L})\gamma^\mu e_L] \nonumber \\
+\frac{1}{2}[P^{11}_\mu \bar{\nu}_L\gamma^\mu \nu_L + P^{12}_\mu \bar{\nu}_L\gamma^\mu e_L \nonumber \\
P^{21}_\mu \bar{e}_L\gamma^\mu \nu_L  + P^{22}_\mu \bar{e}_L\gamma^\mu e_L] 
\end{eqnarray}
\begin{eqnarray}\label{K2}
K_2 = \frac{1}{2}[i  \bar{e}_R\gamma^\mu D_\mu e_R+( i  \bar{e}_R\gamma^\mu D_\mu e_R)^\ast]  \nonumber \\
=\frac{i}{2}  [ \bar{e}_R\gamma^\mu( \partial_\mu e_R)-(\partial_\mu \bar{e_R})\gamma^\mu e_R] \nonumber \\
+\frac{1}{2}[2N\sin^2(\theta_w)Z_\mu-2qA_\mu]\bar{e_R}\gamma^\mu e_R
\end{eqnarray}
The matrix elements $P^{ij}_\mu$  in (\ref{K1}) can be read directly from (\ref{P}).

For the doublet field we have the real invariant form:
\begin{eqnarray}\label{KH}
K_\Phi = (D_\mu\Phi)^\dag D_\mu \Phi =(\partial^\mu\Phi^\ast_i )\partial_\mu\Phi^i \hspace{1.6cm} \nonumber \\ +\frac{i}{2}[\Phi_i^\ast X^{ij}_\mu \partial^\mu \Phi_j -(\partial_\mu \Phi_i^\ast) X_{ij}^\mu \Phi_j] +\frac{1}{4}\Phi_i^\ast X_{ij}^\mu X^{jk}_\mu \Phi_k
\end{eqnarray}  
With the Yang-Mills tensors given by (\ref{YFT}), the Yang-Mills  field Lagrangian in the above notation becomes
\begin{eqnarray}\label{KF}
K_F = -\frac{1}{4}(B_{\mu\nu} B^{\mu\nu} +W_{\mu\nu}^k W^{\mu\nu}_k) = \nonumber \\ 
-\frac{1}{4}( B_{\mu\nu} B^{\mu\nu} +W_{\mu\nu}^3 W^{\mu\nu}_3+ 2 W_{\mu\nu}^+ W^{\mu\nu}_-).
\end{eqnarray}
The components in this expression are given in terms of the electromagnetic potential and boson fields as:
\begin{eqnarray}
B_{\mu\nu} = -\sin(\theta_w) Z_{\mu\nu}+\cos(\theta_w) F_{\mu\nu},W_{\mu\nu}^3 = \nonumber \\ \cos(\theta_w) Z_{\mu\nu}+\sin(\theta_w) F_{\mu\nu} + i g (W_\mu^+ W_\nu^- -W_\mu^- W_\nu^+)
\end{eqnarray}
\begin{eqnarray}
 (W_\mu^1 W_\nu^2 -W_\mu^2 W_\nu^1)  = i(W_\mu^+ W_\nu^- -W_\mu^- W_\nu^+)
\end{eqnarray}
\begin{eqnarray}\label{DW}
W_{\mu\nu}^{\pm}=\frac{1}{\sqrt{2}}(W_{\mu\nu}^1 \pm i W_{\mu\nu}^2) = D^{\pm}_\mu W_\nu^{\pm}- D^{\pm}_\nu W_\mu^{\pm} . 
\end{eqnarray}
Here, $B_{\mu\nu}=\partial_\mu B_\nu  - \partial_\nu B_\mu$, $F_{\mu\nu}=\partial_\mu A_\nu  - \partial_\nu A_\mu$, $Z_{\mu\nu}=\partial_\mu Z_\nu  - \partial_\nu Z_\mu$ and:
\begin{equation}\label{DD}
D_\mu^{\pm} = (\partial_\mu \pm i gW_\mu^3) =  (\partial_\mu \pm i g(\frac{g}{N}Z_\mu+\frac{g^{\prime}}{N}A_\mu))
\end{equation}
Comparison with the electron components above shows that $W^+_\mu$ has the same charge $q$ as the electron, which is negative.

When $\Phi_1 = 0$  and $\Phi_2 \to \nu_o/\sqrt{2}$, the invariant $K_\Phi$ given in (\ref{KH}) becomes
\begin{eqnarray}\label{KHS}
K_\Phi= +\partial^\mu \Phi_2^\ast\partial_\mu\Phi_2-\frac{i}{2}N Z^\mu(\Phi_2^\ast\partial_\mu\Phi_2 - ( \partial_\mu\Phi_2^\ast ) \Phi_2) \nonumber \\ +\frac{1}{4}\Phi_2^\ast\Phi_2(2 g^2 W^- W^+ +N^2 Z^\mu Z_\mu).
\end{eqnarray}
Comparing the latter in the limit $\Phi_2 \to \nu_o/\sqrt{2}$ with the kinetic terms in (\ref{KF}) leads to the following identification for the boson masses. 
\begin{equation}\label{MB}
M_W = \frac{\nu_og}{2}, \hspace{.5cm} M_Z = \frac{\nu_o N}{2}, \hspace{.5cm} M_Z=\cos(\theta_w)M_W
\end{equation}
In this limit we see from (\ref{K1}), (\ref{K2}) and (\ref{K3}), that the electron mass is identified by
\begin{equation}\label{ME}
M_e = G_e\frac{\nu_o}{\sqrt{2}}
\end{equation}

If we combine the invariant expressions in (\ref{K1}), (\ref{K2}), (\ref{K3}), (\ref{KH}) and (\ref{KF}) with a $V(\Phi^4)$ potential, we have the Lagrangian common to the standard electroweak gauge model discussed in numerous text books. 
There are differences at the transformation level from gauge theory.  The nonlinear component of the transformations of the fields depend on the $\xi$ and $h$ field components which do not appear explicitly in the above Lagrangian structure. These fields do appear in the conserved currents associated with the parameters of $G_1$.  Details of the ingredients that go into the conserved currents is provided in the Appendix.

\subsection{Alternate Electroweak Lagrangian}
There are two goals of this section. The first is to describe $SU(2)$  invariant Lagrangian forms involving only the potentials and components of the adjoint field $V^k=Vh^k$. The second goal is to describe a modified invariant Lagrangian for the lepton part. This form used in conjunction with a covariant eigenvalue constraint on the R (right-handed) lepton pair provides an explanation of the absence of a free right-handed neutrino ( $\nu_R$ ) in weak decay. The hypersphere in the adjoint field $V$ becomes a real sphere. 

Consider first the $SU(2)$ invariant vector field $F_\mu$ with the parameter change $(d,\gamma) \to (-g^\prime, g) $.
\begin{eqnarray}
F_\mu = gW_\mu^lV^l -g^{\prime}VB_\mu
\end{eqnarray}
At the poles on the adjoint sphere we have
\begin{eqnarray}
V_3 \to  +V^n: \hspace{.3cm}F^n_\mu = V^nNZ^n_\mu , \hspace{2.5cm} \nonumber \\\
V_3 \to  -V^s: \hspace{.3cm}F^s_\mu = - V^s[Ncos{2\theta_w}Z^s_\mu+2qA^s_\mu].
\end{eqnarray}
The labels $(n,s)$ indicate that the values correspond to the respective poles. We have $F_\mu^n = F_\mu^s $ if the points are not disjoint such as having different vector polarizations. Vector polarization is conserved on the sphere. For instance, if $F_2^s =0$ for a given configuration type, then $F_2=0$ at any other point on the sphere for this configuration type. The massless vector field  $A_\mu$ can exist only at the north pole. 

We consider the two general invariants (\ref{Ka}) and (\ref{Kb}) for $G_1=SU(2)$.
\begin{eqnarray}\label{K5}
K_a = \big[\frac{g^2}{2}\big(V^2W^l_\mu W_l^\mu - W_\mu^l V^lW^\mu_k V^k \big) - gW_\mu^lC_l^\mu \big], \\\
K_b = \frac{1}{2} F_\mu F^\mu \hspace{3cm}
\end{eqnarray}
 In the limit that $V_1 \to 0,V_2 \to 0$ the expression involving the $A^\mu A_\mu$  factor vanishes and the invariant $K_a+K_b$ reduces to
\begin{eqnarray}
 K_a+K_b\to \frac{V^2}{2}(2 g^2 W_\mu^- W_+^\mu +N^2 Z^\mu Z_\mu ).
\end{eqnarray}
The space-time dependence of $V$ is not specified but if we make the notation change $V\to \nu_0 /2$ , we obtain the same form as the expression involving the $W_\mu^{\pm} $ boson and $Z_\mu$ mass terms seen in (\ref{KHS}). With the above potential Lagrangian, the intermediate bosons obtain their masses from the adjoint field $V$. This is a basic difference from the scalar field term  (\ref{KHS}) where these masses arise from the fundamental doublet field. Here, the intermediate boson masses are obtained from a $SU(2)$ vector in contrast to the usual Higgs $SU(2)$ doublet field. This is an important point in light of the fact that the Higgs doublet has not been discovered  even with enormous effort to find it. This shifts the focus from the Higgs doublet to the adjoint, or vector field  as a source of the boson masses. We need to address the possible physical interpretation of the $V$ field. 

One of the first questions is to ask whether or not some components of the  $V$ field are charged. This information must be contained in the Lagrangian.  Consider the following Lagrangian combination for the non lepton part.
\begin{eqnarray}\label{KNL}
K_{NL} = \frac{1}{2}\partial^\mu V^k\partial_\mu V^k + K_F +K_a +K_b = \frac{1}{2}\partial^\mu V^3\partial_\mu V^3 \nonumber \\\ 
 +K_F +(\partial^\mu V^++i g W_3^\mu V^+)(\partial_\mu V^- -i g W_\mu^3 V^-) \hspace{1cm}\nonumber \\\ + \frac{g^2}{2}\Big(2V^2 W^+_\mu W_\mu^-  \hspace{1cm} \nonumber \\\  - (W_\mu^+V^- +W_\mu^-V^+)(W^\mu_+V_- + W^\mu_-V_+)\Big) \hspace{1cm} \nonumber \\\ -g(W_\mu^1C_1^\mu +W_\mu^2C_2^\mu) +K_b \hspace{1cm}
\end{eqnarray}
This expression involves the quantities
\begin{eqnarray}
C_\mu^3 = V_2\partial_\mu V_1-V_1\partial_\mu V_2 = -i(V^+\partial_\mu V^- -V^-\partial_\mu V^+)\nonumber \\\  
V_\bot^2= V_1^2 +V_2^2 = 2V^+V^- \hspace{3cm}.
\end{eqnarray}
Inspection of (\ref{KNL}) shows that the $V^{\pm}$ involve the same covariant derivatives that appear in the field terms for the $W^{\pm}$ potentials. An essential part of this covariant derivative form is the first order kinetic term $ gW_\mu^lC_l^\mu$ required in order for $ K_a$ to be invariant.
The group action  on the $V$ space corresponds to the linear rotation group on the sphere with radius $V$. The action on the potentials is nonlinear. Consider the mass terms $M_A^2A_\mu A^\mu/2$ and $M_Z^2Z_\mu Z^\mu/2$ that occur in the above Lagrangian. With a little effort, we obtain
\begin{eqnarray}
M_A^2 = 2q^2(V^2-VV_3), \nonumber \\\ M_Z^2 = \frac{g^4+(g^{\prime})^4}{N^2}V^2 +2q^2VV_3.
\end{eqnarray}
From these expressions,we obtain the following relation.
\begin{eqnarray}
M_A^2 + M_Z^2 = N^2V^2
\end{eqnarray}
Consider features corresponding to places on the $V$ sphere. At the point $V_3 = V$ we have $M_A=0$ so that this point corresponds to the massless photon field and the intermediate bosons take on the mass values discussed above. However, at any other point masses for the $Z_\mu$ and $A_\mu$ take on different values. It is incorrect to view the $A_\mu$ field as a photon field except at the $V_3 = V$ point. For reference, we call this point the photon pole, or north pole. At $V_3 \to -V$ we have $M_A \to 2qV$. We have the ratio
\begin{eqnarray}
\frac{M_A(V=-V)}{M_Z(V_3=V)} =\frac{2qV}{NV} = \frac{2q}{N}.
\end{eqnarray}
Within the uncertainties of the experimental constants, this number is less than one, but still large so that the $A_\mu$ at the south $V$ space pole is very massive. The $Z_\mu$ particle at this south pole is also massive. Putting in values $g=0.625$, $g^{\prime}=0.357$ and $V = 126.7 Gev/c^2$ we obtain $M_A \simeq 78.5 Gev/c^2$  and $ M_Z \simeq 46.3 Gev/c^2$ at this pole. Small changes in these parameters will not change the fact that these bosons are very heavy at this pole. With that said, the interaction terms will produce significant energy shifts from these values in a more complete solution. However, except at the north $V$ space pole, we have two massive interacting "neutral" fields $A_\mu$ and $Z_\mu$. 

The form in the standard model for the right-handed electron invariant (\ref{K2}) permitted the $SU(2)$ symmetry while accommodating (but not explaining) the absence of the right-handed neutrino in in weak interactions. Here, we consider an alternate picture that offers an explanation of the absence of the right-handed neutrino in such processes. We use the notation $ \nu, e $ for two-four component spinors that will become the neutrino and electron respectively only in certain limits, but will take on different interpretations in other cases. 
To construct an invariant lepton term, we consider an alternate formulation for which $Ye_R = -1e_R$ with the right-handed field $ \mathbf{R}$ defined by 
\begin{equation}
\mathbf{R} = \left(\begin{array}{c}\nu_R\\e_R\end{array}\right),   \tilde{\mathbf{R}} = \left(\begin{array}{cc} \bar{\nu_R}& \bar{e_R}\end{array}\right). 
\end{equation}
The transformation on $ \mathbf{R}$ and corresponding covariant derivative are exactly like those for $\mathbf{L} $. 
\begin{eqnarray}\label{TRs}
[T_a,\mathbf{R}]     =  \frac{i}{2}\mathbf{\sigma_a} \mathbf{R} + i\xi_a \mathbf{\mathcal{T^-} }\mathbf{R} 
\end{eqnarray}
\begin{equation}\label{DR:2}
D_\mu \mathbf{R} = \partial_\mu \mathbf{R}   - \frac{i}{2}\mathbf{P}\mathbf{R} 
\end{equation}
We replace (\ref{K2}) with the following invariant 
\begin{eqnarray}\label{K2M}
KM_2= \frac{1}{2}[i  \tilde{\mathbf{R}}\mathbf{\Gamma}^\mu D_\mu \mathbf{R}  +(i  \tilde{\mathbf{R}}\mathbf{\Gamma}^\mu D_\mu \mathbf{R} )^\ast] \nonumber \\ =\frac{i}{2}[\bar{\nu}_R\gamma^\mu( \partial_\mu \nu_R)-(\partial_\mu \bar{\nu_R})\gamma^\mu \nu_R]  \nonumber \\ +\bar{e}_R\gamma^\mu( \partial_\mu e_R)-(\partial_\mu \bar{e_R})\gamma^\mu e_R] \nonumber \\
+\frac{1}{2}[P^{11}_\mu \bar{\nu}_R\gamma^\mu \nu_R + P^{12}_\mu \bar{\nu}_L\gamma^\mu e_R \nonumber \\
P^{21}_\mu \bar{e}_R\gamma^\mu \nu_R  + P^{22}_\mu \bar{e}_R\gamma^\mu e_R] 
\end{eqnarray}
The invariant (\ref{K3}) is replaced by the invariant
\begin{eqnarray}\label{KM3}
KM_3 = -m [ \tilde{\mathbf{L}}\mathbf{R} + \tilde{\mathbf{R}}\mathbf{L} ] \nonumber \\  = -m[\bar{\nu_R}\nu_L +  \bar{\nu_L} \nu_R]  -m[\bar{e_R}e_L +  \bar{e_L} e_R ].
\end{eqnarray}
Here $m$ is the mass of the electron. The reader should recall that in the standard model the electron mass term involved a product of a constant times a Higgs doublet component. Here, we could express the mass as a product of a constant times the invariant $V$, but we have no symmetry justification for it. Both bracket expressions in (\ref{KM3}) are needed for $SU(2)$ invariance. 

We now turn our attention to the question about the right-handed neutrino $\nu_R$. We impose the following  matrix eigenvalue constraint on $\mathbf{R}$.
\begin{equation}\label{C}
H\mathbf{R} = \left[\begin{array}{cc}h_3 & (h_1-i h_2) \\  (h_1+i h_2) & -h_3 \end{array}\right] \mathbf{R} = \lambda\mathbf{R}.
\end{equation}
The eigenvalues are
\begin{equation}
\lambda_{\pm} = \pm h, \hspace{.5cm}  h = \sqrt{h^k h^k} = 1.
\end{equation}
This matrix eigenvalue equation is covariant under the group for either of the eigenvalues. 
Here, we consider the case for $\lambda = -1$, for which the two dimensional eigenvalue constraint can be expressed by the following two equations.
\begin{eqnarray}\label{Eig}
(1+h_3)\nu_R + \sqrt{2}h^-e_R= 0, \\\ \sqrt{2}h^+\nu_R + (1-h_3)e_R = 0
\end{eqnarray}
These two equations require that the right-handed neutrino $\nu_R$ vanish at the north pole $h_3=1$. This is the point where the $A_\mu$ field becomes massless, and the intermediate boson masses take on the observed relative ratios. This is one reasonable explanation of the observed absence of the right-handed neutrino in weak interactions. This constraint does not mean that  $\nu_R$ vanish at other points on the $h$ sphere. At points other than at $h_3=-1$ we may re-express the first of these constraint equations as
\begin{equation}
\nu_R = \frac{-\sqrt{2}h^-}{1+h_3}e_R.
\end{equation}
What does this mean? It means that at places other than the $V_3=V$ pole the spinor $\nu$ has become massive, and is thus no longer a neutrino. With this, consider that the constraint (\ref{C}) requires that the $e_R$ spinor vanish at the south pole $V_3=-V$, so that the spinor $e$ becomes massless. Is it a left-handed neutrino? The answer is no, unless all of the massive potentials including $A_\mu$ vanish. In this case,  it becomes a different type of left handed neutrino as determined by the way it interacts. With the presence of a nonzero $A_\mu$ field in the Lagrangian, one is tempted to think that this massless spinor would be affected by the electromagnetic field. This is not the case because at this south pole the $A_\mu$ field is very massive, and perhaps short lived. The $A_\mu$ field like the spinor fields morph between massive and massless fields with changing positions on this sphere. At this $V_3=-V$ pole, the spinor $\nu$ becomes massive, and is again no longer a neutrino. With the above Lagrangian, the $\nu$ spinor has the mass of the electron at the $V_3=V$ pole.

Consider a a view of the whole sphere.  Points other than the north and south pole correspond to nonzero $V^{\pm}$ fields.  Recall from (\ref{DW}), (\ref{DD}) and (\ref{KNL}) that the $V^{\pm}$ fields and the $W^{\pm}_\mu$ fields have the same covariant derivative 
\begin{eqnarray}
D_\mu^{\pm} = (\partial_\mu \pm i gW_\mu^3) =  (\partial_\mu \pm i g(\frac{g}{N}Z_\mu+\frac{g^{\prime}}{N}A_\mu)).
\end{eqnarray}
At points other than the north pole, both the $A_\mu$ and $Z_\mu$ fields are massive. It would be impossible to manipulate the $V{\pm}$ and $W^{\pm}_\mu$ fields with an apparatus based on a massless field at points other than the north pole. There is no interaction with the massless field except at the north pole. There, the $V^{\pm}$ fields vanish. There, we should observe the $W^{\pm}_\mu$ fields, and we do. We could observe the $V^{\pm}$ fields if we could build an apparatus based on the massive $A_\mu$ field. What about the leptons at places other than the poles? The two leptons are massive, and interact with the massive potential fields. With the constraint (\ref{C}) the two right-handed components are coupled, so that in deriving the field equations from the Lagrangian, this constraint must be invoked. Because of this constraint, the leptons will interact with the $V$ field even if the massive potentials vanish.

The question is "What physical interpretation do these massive leptons have at non-pole points?" They have mass and should play a role in gravity, especially if there are many of them. They are difficult to detect by usual means. They have the characteristics of WIMPS, offering a  Fermion contribution for part of the "missing" mass. The $V^{\pm}$ fields and the four massive potentials offer a boson explanation for part of the missing mass. In fact, any massive field or potential in the non-pole part of the $V$ sphere could offer a contribution to the missing mass. This non-pole world is observable by gravity, otherwise, it is difficult to observe because there we do not have the massless electromagnetic field to help in acceleration and detection of these particles. Put in layman language, these particles do not see the massless electromagnetic field that exist only at the north pole. 

In the above we have considered the constraint for $\lambda=-1$ on the right leptons. For the $\lambda=+1$ eigenvalue we have the following constraint equations.
\begin{eqnarray}\label{Eig2}
(-1+h_3)\nu_R + \sqrt{2}h^-e_R= 0, \\\ \sqrt{2}h^+\nu_R - (1+h_3)e_R = 0
\end{eqnarray}
From this we see that $e_R \to 0$ at the north pole but there is no restriction on the right-handed neutrino. The neutrino can have a mass if the right handed neutrino does not vanish there for some other reason. At the south pole, the right-handed neutrino must vanish, and the "electron" spinor remains massive. Its interpretation is not clear since there is no massless (electromagnetic) field. For either eigenvalue case there is no massive left-handed neutrino when the right-handed neutrino vanishes. The invariants $K_a$ and $K_b$ in this section are independent of whether or not we impose constraints on the right handed leptons. 

\subsection{Summary}

 In this study we have described in some detail the Lie Algebra of a particular type of nonlinear realization of Lie groups that leads to interactions via covariant derivatives.  These realizations differ from local gauge realizations in that the group parameters are global, and the interaction arises from the dependence, generally nonlinear, of the transformation of one field on another.  These realizations are characterized by a field $V$ whose components transforms via the adjoint representation. These can be viewed as linear transformations on a hypersphere with radius given by $V = \sqrt{V^kV^k}$. The surface of this hypersphere provides a convenient  reference to describe the changes of the fields and potentials. 
 
In section \textbf{B} we presented two invariant forms involving the potentials and $V$ space components but not the leptons.   Covariant matrix constraints, which if imposed, place restrictions on certain field components so that they vanish at certain places on the hypersphere. Fields, potentials and features morph in changing from one position to another. For instance, masses can change, going in some cases from very massive to massless. Different positions on the hypersphere at a given space-time point correspond to the presence of different fields and potentials at this space-time point. The physical interpretation of the symmetry is active.
 
The above general features are made more specific by a study of the electroweak interaction in this picture. In section \textbf{C} we first show that the standard gauge electroweak Lagrangian is invariant in this picture. As in gauge theory this invariant form makes use of a Higgs doublet to build around the fact that the right-handed neutrino is not observed. 

In section \textbf{D}, we presented an alternate Lagrangian structure together with a covariant constraint on the right- handed lepton field. We use the same constants as used in the standard model described in the previous section.The hypersphere becomes a three dimensional sphere. In different surface zones on this sphere, the physical fields differ to the extent that fields in one zone can have different features than fields in another zone. The first zone of interest is the north pole $(V^3=V)$. There we obtain a collection of events that happens no where else on the sphere. First, the $A_\mu$ field becomes massless. At this point we can have electromagnetic interactions. An invariant lepton mass term is used in which the right-handed neutrino field is not assumed \textit{a priori} to be zero. An eigenvalue constraint on the right-handed lepton pair is invoked. It requires that the right-handed neutrino to vanish, but only at this point. The last thing that happens at this point is that the alternate invariant Lagrangian at this point provides the same boson mass ratio $\frac{M_Z}{M_W} $ found in the standard gauge model. This is the only point where this happens.These features are consistent with observation. The mass scale of the bosons is determined by the radius $V$ of the adjoint field, not the Higgs doublet. It would appear that a great deal of our observable physics happens at the north pole. Are other points on the sphere hidden from observation except via gravity? Perhaps not, but for these points, observations need to be done, or classified via non electromagnet means.

A different zone is the south pole $(V_3=-V)$. There, the $A_\mu$ field is very massive and we have no massless potential. The covariant constraint with $\lambda=-1$ requires that the right-handed electron vanish at this point. The left-handed electron becomes massless. The physical interpretation is not clear. Since there is no massless field here to represent the electromagnetic field, we could ask how this "electron" could be observed. At this south pole for this constraint, the lepton component that was the neutrino at the north pole now has the mass of the electron, but it is not an electron.

At all  zones on the sphere other than the two poles both lepton components have light masses. All four boson potential fields are very masses. The mass of the $A_\mu$ field decreases to zero as one approaches the north pole, but is very heavy near the south pole.  This large zone between the poles may be difficult to access in the laboratory because in this zone we have no massless field. Fields in this zone are blind to the electromagnetic field. Because of the large masses, interactions would perhaps be fast. The two low mass leptons and four heavy boson fields in this zone between the poles offer possible support for a WIMP contribution to the missing mass. 

It is incorrect to call the leptons, $V$ space components and vector bosons at points other than the north pole "dark matter". They simply cannot be seen with electromagnetic eyes that exist only at the north pole in this model. Observation will depend on appropriate detectors for this region, just like different detectors are needed for different regions of the electromagnetic spectrum. The existence of different types of neutrinos has been observed. Some "neutrinos" already detected by different types of neutrino detectors may, in reality, be the light mass leptons at points other than the north pole on the $V$ sphere. Just what is the signature of a neutrino, or more directly, when is a lepton a neutrino?

The realizations presented here provide a conserved current for the linear part and a conserved current for the nonlinear.  The form of the latter current corresponds to conservation of charge at the north pole. These currents are described in detail in the \textbf{Appendix}. The general development in sections \textbf{A} and \textbf{B} was presented as a foundation not only for the $SU(2)$ electroweak application here but for possible future applications of $SU(3)$, and other Lie groups. Two begging question raised by this study are:  Can an explanation of quark confinement be obtained via this approach? Do "free" quarks exist but are just blind to our usual means of detection?
 
\subsection{Appendix: Conserved Currents}

Here, we first look at the details of the components that go into the conserved currents for the standard Lagrangian made from a sum of the  (\ref{K1}), (\ref{K2}), (\ref{K3}), (\ref{KH}) and (\ref{KF}) terms together with a $V(\Phi^4)$ potential. We then provide details for the conserved currents for the modified Lagrangian. With a real Lagrangian we can use $J^\mu_a(F_i^\ast) = J^\mu_a(F_i)^\ast $ for each complex field component $F_i$. The generator action for each field and potential separates into a linear and nonlinear part. 
\begin{eqnarray}
[T_a, F_i] = [T_a, F_i]_L  + [T_a, F_i]_{NL} 
\end{eqnarray}
The "L" label for linear should not be confused with the same label for left on the spinors. Both the linear and nonlinear current parts are separately conserved since when $\xi_a\to 0$ the nonlinear current vanish and the linear parts are unchanged. Separate conserved currents were also obtained in the nonlinear extension of representations of $SL(2,C)$ in \cite{dc}.
 Below, we list the  detailed expressions for both factors in the following current terms for each variable in the Lagrangian. 
\begin{eqnarray}
J^\mu_a(F_i) =\Pi^\mu(F_i)[T_a,F_i], \hspace{.3cm} \Pi^\mu(F_i) = \frac{\partial K}{\partial F_{i,\mu}},  \nonumber \\\
\end{eqnarray}
To facilitate constructing the conserved linear and conserved nonlinear currents, we separate the linear and nonlinear generator action in the following list. For the spinors we have the forms:
\begin{eqnarray}
\Pi^\mu (\nu_L) = \frac{i}{2} \bar{\nu_L} \gamma^\mu \hspace{3cm} \hspace{2cm} \nonumber \\\  [T_a,\nu_L]_L = \frac{i}{2}(\sigma_a^{11}\nu_L + \sigma_a^{12}e_L), \hspace{3cm} \nonumber \\\
[T_a , \nu_L]_{NL} = \frac{i}{2} \xi_a [(h_3-1)\nu_L +\sqrt{2} h^- e_L] \hspace{2cm}
\end{eqnarray}
\begin{eqnarray}
 \Pi^\mu (e_L) = \frac{i}{2} \bar{e_L} \gamma^\mu\hspace{3cm} \hspace{2cm}  \nonumber \\\  [T_a,e_L]_L = \frac{i}{2}(\sigma_a^{21}\nu_L + \sigma_a^{22}e_L),  \hspace{3cm} \nonumber \\\
[T_a , e_L]_{NL} = \frac{i}{2} \xi_a [\sqrt{2} h^+ \nu_L - (h_3+1)e_L] \hspace{2cm}
\end{eqnarray}
\begin{eqnarray}
\Pi^\mu (e_R) = \frac{i}{2} \bar{e_R} \gamma^\mu, \hspace{.5cm} [T_a , e_R] = - i \xi_a e_R \hspace{1cm}
\end{eqnarray}
There is no linear action on the singlet $e_R$ field. For the alternate picture, the right-handed lepton factors are like those factors above for the left-handed leptons.
For the potentials, we have:
\begin{eqnarray}
\Pi^\mu (W^{\pm}_\rho) = - W_{\mp}^{\mu\rho}\hspace{4cm} \nonumber \\\   [T_a , W_\rho^{\pm}]_L = \frac{1}{\sqrt{2}}(\epsilon^{a1k} \pm i\epsilon^{a2k})W_\rho^k \hspace{1.5cm}  \nonumber \\\
[T_a , W_\rho^{\pm}]_{NL} = \pm i \xi_a h^{\pm}[\cos(\theta_w) Z_\rho+\sin(\theta_w) A_\rho ] \nonumber \\\
\mp i \xi_a W_\rho^{\pm} h_3+\frac{1}{g}\partial_\rho (\xi_a h^{\pm}) 
\end{eqnarray} 
\begin{eqnarray}
\Pi^\mu(Z_\rho) = -Z^{\mu\rho}+iN\cos^2(\theta_w)[W^\mu_+W^\rho_--W^\rho_+W^\mu_-]  \nonumber \\\ [T_a, Z_\rho]_L=\cos(\theta_w) \epsilon^{a3k}W_\rho^k \hspace{3cm}\nonumber \\\
[T_a, Z_\rho]_{NL} = -i \xi_a \cos(\theta_w) [h^+ W^-_\rho -h^-W^+_\rho] \hspace{.7cm}\nonumber \\\ +\frac{1}{N} [\xi_a\partial_\rho h_3 + (h_3-1)\partial_\rho\xi_a] \hspace{2cm}
\end{eqnarray}
\begin{eqnarray}
\Pi^\mu(A_\rho) = -F^{\mu\rho}+iq[W^\mu_+W^\rho_--W^\rho_+W^\mu_-]\hspace{1cm} \nonumber \\\ [T_a, A_\rho]_L=\sin(\theta_w) \epsilon^{a3k}W_\rho^k \hspace{3cm} \nonumber \\\
[T_a, A_\rho]_{NL} = -i \xi_a \sin(\theta_w) [h^+ W^-_\rho -h^-W^+_\rho] \hspace{1cm}\nonumber \\\    +\frac{\tan(\theta_w)}{N} \xi_a\partial_\rho h_3    +\frac{1}{q} [\sin^2(\theta_w) h_3 +\cos^2(\theta_w)]\partial_\rho\xi_a
\end{eqnarray}
Finally, for the scalar components we have:
\begin{eqnarray}
\Pi^\mu (\Phi_1)  = \partial^\mu \Phi_1^\ast +\frac{i}{2}\Big([ N\cos{2\theta_w} Z_\mu + 2qA_\mu]\Phi_1^\ast  \hspace{.3cm} \nonumber \\\ +g \sqrt{2}W_+^\mu\Phi_2^\ast \Big)\hspace{1cm} \nonumber \\\ [T_a,\Phi_1]_L = \frac{i}{2}(\sigma_a^{11}\Phi_1 + \sigma_a^{12}\Phi_2), \hspace{3cm}  \nonumber \\\
[T_a,\Phi_1]_{NL} =   \frac{i}{2}\xi_a [(1+h_3)\Phi_1 +\sqrt{2}h^-\Phi_2] \hspace{1cm}
\end{eqnarray}
\begin{eqnarray}
\Pi^\mu (\Phi_2)  = \partial^\mu \Phi_2^\ast +\frac{i}{2}\xi_a  \big[ g\sqrt{2}W_-^\mu\Phi_1^\ast  - NZ^\mu\Phi_2^\ast \big] \hspace{.7cm} \nonumber \\\  [T_a,\Phi_2]_L = \frac{i}{2}(\sigma_a^{21}\Phi_1 + \sigma_a^{22}\Phi_2), \hspace{2.5cm} \nonumber \\\
[T_a,\Phi_2]_{NL} = \frac{i}{2}\xi_a [\sqrt{2}h^+\Phi_1 +(1-h_3)\Phi_2] \hspace{1cm}
\end{eqnarray}

Each variable in the Lagrangian contributes to the conserved current for each parameter of $G_1$. The linear current is obtained by taking products of the term and adding the complex conjugate where appropriate. The nonlinear part needs some discussion. We consider the nonlinear part below, leaving off the NL label for convenience.
The lepton contribution is
\begin{eqnarray}\label{Jlep}
J^\mu_a(l) = \frac{\xi_a }{2}\Big[(1-h_3)\bar{\nu_L}\gamma_\mu \nu_L  + (1+h_3)\bar{e_L}\gamma_\mu e_L \nonumber \\\  -\sqrt{2}h^-\bar{\nu_L}\gamma_\mu e_L- \sqrt{2}h^+\bar{e_L}\gamma_\mu \nu_L \Big] +\xi_a \bar{e_R}\gamma_\mu e_R. \hspace{.5cm}
\end{eqnarray}
The individual potential contributions to the conserved currents are
\begin{eqnarray}
J^\mu_a(W_+) = -W^{\mu\rho}_-\Big[- i \xi_a h^+[\cos(\theta_w) Z_\rho+\sin(\theta_w) A_\rho ] \nonumber \\\
+ i \xi_a W_\rho^+ h_3+\frac{1}{g}\partial_\rho (\xi_a h^+) \Big], \hspace{1cm} \nonumber \\\
J^\mu_a(W_-) = -W^{\mu\rho}_+\Big[+ i \xi_a h^-[\cos(\theta_w) Z_\rho+\sin(\theta_w) A_\rho ] \nonumber \\\
- i \xi_a W_\rho^- h_3+\frac{1}{g}\partial_\rho (\xi_a h^-) \Big] \hspace{1cm}, \nonumber \\\
J^\mu_a(Z) = \Big(-Z^{\mu\rho}+iN\cos^2(\theta_w)[W^\mu_+W^\rho_--W^\rho_+W^\mu_-] \Big) \nonumber \\\ \Big( i \xi_a \cos(\theta_w) [h^+ W^-_\rho -h^-W^+_\rho] \hspace{.7cm}\nonumber \\\ +\frac{1}{N} [\xi_a\partial_\rho h_3 + (h_3-1)\partial_\rho\xi_a] \Big), \nonumber \\\
J^\mu_a(A) = \Big(-F^{\mu\rho}+iq[W^\mu_+W^\rho_--W^\rho_+W^\mu_-] \Big) \hspace{1.5cm} \nonumber \\\
\Big( i \xi_a \sin(\theta_w) [h^+ W^-_\rho -h^-W^+_\rho] +\frac{\tan(\theta_w)}{N} \xi_a\partial_\rho h_3  \nonumber \\\  +\frac{1}{q} [\sin^2(\theta_w) h_3 +\cos^2(\theta_w)]\partial_\rho\xi_a \Big). \hspace{1cm}
\end{eqnarray}
For the scalar field we have the following current contribution.
\begin{eqnarray}
J^\mu_a(\Phi)=i\xi_a\big[(\partial^\mu\Phi^\dagger) T^+\Phi-\Phi^\dagger T^+\partial^\mu \Phi\big] \nonumber \\\
-\frac{\xi_a}{4} \big[\Phi^\dagger( X^\mu T^+ + T^+X^\mu)\Phi\big]
\end{eqnarray}
For use with the alternate Lagrangian the above lepton contribution is replaced by 
\begin{eqnarray}\label{Rep}
J^\mu_a(l) = \frac{\xi_a }{2}\Big[(1-h_3)\bar{\nu_L}\gamma_\mu \nu_L  + (1+h_3)\bar{e_L}\gamma_\mu e_L \nonumber \\\  -\sqrt{2}h^-\bar{\nu_L}\gamma_\mu e_L- \sqrt{2}h^+\bar{e_L}\gamma_\mu \nu_L \Big]  \nonumber \\\  + \frac{\xi_a }{2}\Big[(1-h_3)\bar{\nu_R}\gamma_\mu \nu_R  + (1+h_3)\bar{e_R}\gamma_\mu e_R \nonumber \\\  -\sqrt{2}h^-\bar{\nu_R}\gamma_\mu e_R- \sqrt{2}h^+\bar{e_R}\gamma_\mu \nu_R \Big] .
\end{eqnarray}

By summing the above terms, we obtain a separate conserved current for each group parameter. However the terms include  $\xi_a$ and $\partial^\mu \xi_a$ factors differ for each group parameter and do not appear in the Lagrangian. The above total current has the general form
\begin{eqnarray}
J^\mu_a = \xi_aC^\mu +\Sigma_iS^{\mu\rho}_iF_\rho^i\xi_a + \Sigma_iS^{\mu\rho}_iG^i\partial_\rho\xi_a.
\end{eqnarray}
Here, first term is for all non potential parts, and the last two terms represent the potential contributions. Here, $ S^{\mu\rho}_i=-S^{\rho\mu}_i$ and the sum is over the contributions from the four potentials. Using (\ref{cc}), we have 
\begin{eqnarray}
(\partial_\mu\xi_a)\big[C^\mu + \Sigma_iS^{\mu\rho}_iF_\rho^i + \Sigma_i\partial_\rho(S^{\rho\mu}_iG^i)\big] \nonumber \\
+\xi_a\partial_u\big[C^\mu + \Sigma_iS^{\mu\rho}_iF_\rho^i \big] =0.
\end{eqnarray}
Since the $\xi_a$ will generally differ for each group parameter, we satisfy this relation for all group parameters with 
\begin{eqnarray}
C^\mu + \Sigma_iS^{\mu\rho}_iF_\rho^i + \Sigma_i\partial_\rho(S^{\rho\mu}_iG^i) =0, 
\end{eqnarray}
\begin{eqnarray}\label{ccc}
\partial_\mu j^\mu=0, \hspace{.5cm} j^\mu = C^\mu + \Sigma_iS^{\mu\rho}_iF_\rho^i.
\end{eqnarray}
These two relations are mutually consistent since taking $\partial_u$ on the first leads directly to the second because of 
the relation $ S^{\mu\rho}_i=-S^{\rho\mu}_i$.
The last expression means that the little currents $j^\mu$ are conserved. These currents do not involve the $\xi_a$components. What this means is that contained within these nonlinear realizations is a common conserved quantity, independent of any particular group parameter. 

The conserved linear currents can be easily constructed from the above factors. Care must be taken to add the complex conjugate field contribution where appropriate. What is the physical interpretation of the little current $j_\mu$ that arises via the nonlinear transformations? Some insight can be gained by looking at the little current at the two poles $h_3=\pm1$. At the north pole $(h_3=1)$ for the standard Lagrangian, we have from the above factors
\begin{eqnarray}\label{NP}
j^\mu = \bar{e_L}\gamma^\mu e_L + \bar{e_R}\gamma^\mu e_R -i\big[W_-^{\mu\rho}W^+_\rho- W_+^{\mu\rho}W_\rho^-\big]  \nonumber \\\ +i\big[(\partial^\mu \Phi_1^\ast)\Phi_1 - (\partial^\mu \Phi_1)\Phi_1^\ast \big] \hspace{2cm}  \nonumber 
\\\ +\big[N\cos{2\theta_w} Z_\mu + 2qA_\mu\big]\Phi_1^\ast\Phi_1\hspace{2cm} \nonumber \\\ -\frac{g}{\sqrt{2}}\big[W_+^\mu\Phi_2^\ast \Phi_1 + W_-^\mu\Phi_2\Phi_1^\ast \big]. \hspace{2cm}
\end{eqnarray}  
The little current at this pole is proportional to the electromagnetic current density.  
For the alternate Lagrangian we drop the terms involving the Higgs doublet $\Phi$. At the north pole the $V$ space components do not contribute to the little current. At this pole $V^{\pm} = 0$. The usual gauge theory practice is to set $\Phi_1=0$ and $\Phi_2= \frac{\nu_o}{\sqrt{2}}$, a constant, in lowest order to generate the intermediate boson masses with the Higgs doublet. The alternate Lagrangian generates the boson masses via the adjoint $V$ field.

At the south pole $(h_3=-1)$  with the modified  Lagrangian we have
\begin{eqnarray}\label{SP}
j^\mu = \bar{\nu_L}\gamma^\mu \nu_L + \bar{\nu_R}\gamma^\mu \nu_R \hspace{1.5cm}  \nonumber \\\ + i\big[W_-^{\mu\rho}W^+_\rho- W_+^{\mu\rho}W_\rho^- \big].  
\end{eqnarray}
Recall that at this south pole the $\nu$ field has a mass and the $e_L$ field has become massless. At neither pole do the massless lepton fields contribute to the conserved little current. Recall that at the south pole the $A_\mu$  boson field is massive, so interpretation of the little current at this pole is not as clear. Nevertheless, it is conserved. For possible use, the Higgs field at this south pole would contribute the following to the little current.
\begin{eqnarray}
i\big[(\partial^\mu \Phi_2^\ast)\Phi_2 - (\partial^\mu \Phi_2)\Phi_2^\ast \big]   + N Z_\mu \Phi_2^\ast\Phi_2\hspace{1cm} \nonumber \\\ -\frac{g}{\sqrt{2}}\big[W_-^\mu\Phi_1^\ast \Phi_2 + W_+^\mu\Phi_1\Phi_2^\ast \big]. \hspace{2cm}
\end{eqnarray}
Expressions for the currents at places other than the poles are a bit complicated, but if needed, can be obtained directly from the factors discussed above.

\begin{acknowledgments}
The author would like to think Kevin Haglin for reading this manuscript and for many useful discussions on the variety of topics raised in this study.
\end{acknowledgments}

\end{document}